\documentclass[10pt,aps,prc,twocolumn,superscriptaddress,preprintnumbers]{revtex4-1}
\usepackage{tabularx} 
\usepackage{graphicx} 
\usepackage{hyperref}  
\usepackage{amssymb}  
\usepackage{amsmath}  
\newcommand{\be}{\begin{eqnarray}}
\newcommand{\ee}{\end{eqnarray}}
\newcommand{\beq}{\begin{equation}}
\newcommand{\eeq}{\end{equation}}
\begin{document}
\title{Proton charge radius extraction from electron scattering data \\ 
using dispersively improved chiral effective field theory}
\author{J.~M.~Alarc\'{o}n}
\affiliation{Departamento de F\'isica Te\'orica, Universidad Complutense de Madrid, 28040 Madrid, Spain}
\author{D.~W.~Higinbotham}
\affiliation{Jefferson Lab, Newport News, VA 23606}
\author{C.~Weiss}
\affiliation{Jefferson Lab, Newport News, VA 23606}
\author{Zhihong Ye}
\affiliation{Argonne National Lab, Argonne, IL 60439}
\begin{abstract}
We extract the proton charge radius from the elastic form factor (FF) data using a novel theoretical 
framework combining chiral effective field theory and dispersion analysis. 
Complex analyticity in the momentum transfer correlates the behavior of the spacelike FF
at finite $Q^2$ with the derivative at $Q^2 = 0$. The FF calculated in the predictive
theory contains the radius as a free parameter. We determine its value by comparing the predictions 
with a descriptive global fit of the spacelike FF data, taking into account the theoretical
and experimental uncertainties. Our method allows us to use the finite-$Q^2$ FF data 
for constraining the radius (up to $Q^2\sim$ 0.5 GeV$^2$ and larger) 
and avoids the difficulties arising in methods relying on the $Q^2 \rightarrow 0$ extrapolation.
We obtain a radius of 0.844(7) fm, consistent with the high-precision muonic hydrogen results.
\end{abstract}
\preprint{JLAB-THY-18-2804}
\maketitle
\section{Introduction}
The proton charge radius is a fundamental quantity of nuclear physics and attests to the 
hadron's finite spatial extent and composite internal structure. It is defined as 
the derivative of the proton electric form factor (FF) at zero momentum transfer, 
$(r_E^p)^2 \equiv -6 \, d G_E^p/dQ^2 (Q^2 = 0)$, and describes the leading finite-size
effect in the interaction with long-wavelength electric fields; see Ref.~\cite{Miller:2018ybm}
for a critical discussion of its interpretation. 
The electric and magnetic FFs at $Q^2 > 0$ are measured in elastic electron-proton 
scattering experiments; see Refs.~\cite{Punjabi:2015bba,Pacetti:2015iqa} for a review. 
The radius is also extracted from nuclear corrections to atomic energy levels measured in 
precision spectroscopy experiments. Measurements of muonic hydrogen transitions have obtained 
a value $r_E^p =$ 0.84087(39)~fm \cite{Pohl:2010zza,Antognini:1900ns}, significantly smaller than the value of 0.875 fm by the Committee on Data for Science and
Technology (CODATA), obtained from electronic hydrogen transitions and some information 
from electron scattering data \cite{Mohr:2015ccw}. The discrepancy, known as the ``proton radius 
puzzle,'' is the subject of a lively debate and has stimulated extensive theoretical 
and experimental research \cite{Pohl:2013yb,Carlson:2015jba}, including new dedicated
low-$Q^2$ electron-proton and muon-proton scattering experiments \cite{Gasparian:2014rna,Gilman:2017hdr}.

Determining the charge radius from electron scattering data amounts to inferring 
the derivative of the FF at $Q^2 = 0$ from the data at finite $Q^2$.
From an empirical point of view, the problem presents itself as one of ``extrapolation'' of 
the measured FF to $Q^2 \rightarrow 0$. Two approaches have been taken in most studies 
so far; see Ref.~\cite{Yan:2018bez} for a review and \cite{Shmueli:2010} for the general concepts.
Descriptive fits (e.g.\ higher-order polynomial fits) provide excellent descriptions
of the data over a wide range of $Q^2$, but the functions are generally not well-behaved outside
the fitted region \cite{Bernauer:2010wm,Bernauer:2013tpr,Lee:2015jqa}.
Predictive models (e.g.\ fits with low-order polynomials or other smoothly varying 
functions) permit stable extrapolation but are constrained by either the selected functional 
form or tightly bounded 
parameters \cite{Griffioen:2015hta,Higinbotham:2015rja,Horbatsch:2016ilr,Hayward:2018qij,Higinbotham:2018jfh}.
In both approaches the question arises over what $Q^2$ range the extrapolation 
should optimally be performed, and what uncertainties are associated with this choice. 

Complex analyticity plays an essential role in the behavior of the proton FF at low $Q^2$.
The FF is an analytic function of $Q^2$, with singularities at $Q^2 < 0$, starting 
with the two-pion cut at $Q^2 < -4 M_\pi^2$. The behavior of the FF at $Q^2 > 0$,
where it is measured in elastic scattering, is governed by the position of these 
singularities and by their strength (spectral function), which can be calculated using 
theoretical methods. Analyticity thus implies correlations between the behavior of the 
FF in different regions of the $Q^2 > 0$ domain, which are not apparent in purely descriptive fits.
It should therefore inform the analysis of low-$Q^2$ FF data and extraction of the radius
\cite{Hohler:1976ax,Belushkin:2006qa,Lorenz:2012tm}. In this way one can go beyond the method of 
extrapolation and use data in a wider $Q^2$ domain to constrain the derivative at zero.

Here we report an extraction of the proton charge radius using a novel predictive theoretical
framework that implements analyticity -- dispersively improved chiral effective field theory (DI$\chi$EFT)
\cite{Alarcon:2017ivh,Alarcon:2017lhg,Alarcon:2018irp}. We express the spacelike proton FF predicted by the 
theory in a form such it contains the radius as a free parameter, which directly exhibits the 
correlation between the finite-$Q^2$ behavior and the derivative at $Q^2 = 0$ implied by analyticity.
We determine the value of the radius by comparing the theoretical predictions with a descriptive global 
fit of the spacelike FF data \cite{Ye:2017gyb} over a broad range of $Q^2$ (optimally up to 
$\sim 0.5$ GeV$^2$), taking into account the theoretical and experimental uncertainties.
At the ``best'' radius the theory describes the data with the same accuracy as the global fit.
Our approach thus combines the best features of descriptive and predictive modeling.
It recruits the finite-$Q^2$ FF data for constraining the radius and overcomes the theoretical
and experimental limitations of the $Q^2 \rightarrow 0$ extrapolation. We obtain a radius of 
0.844(7)~fm, which reconciles the electron scattering data with the muonic hydrogen value.
We comment on possible improvements of the radius extraction and the relation to other approaches.
%
%
\begin{figure*}[t]
\includegraphics[width=2.0\columnwidth]{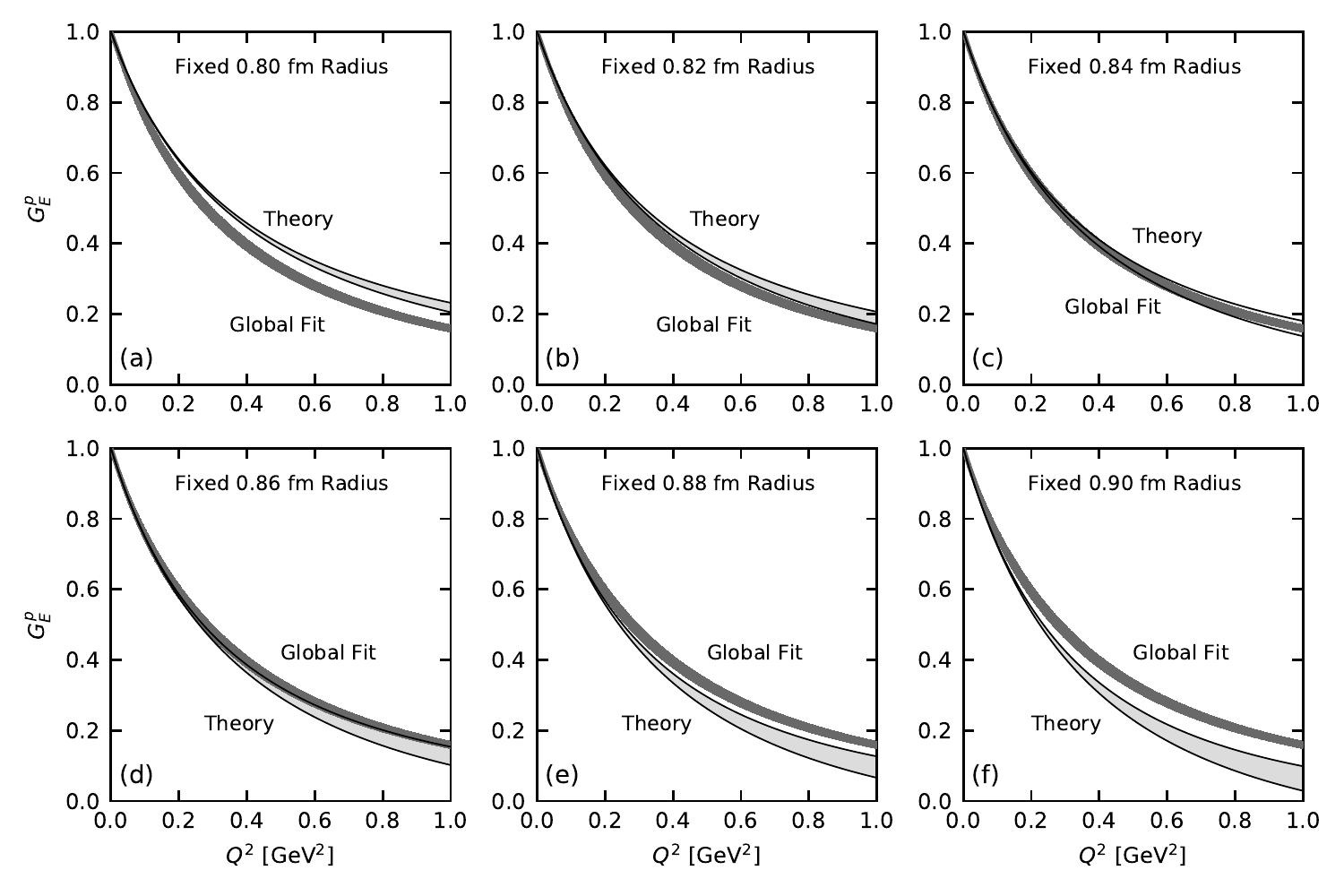}
\caption{{\textit{Light shaded bands between solid lines (labeled ``Theory''):} 
DI$\chi$EFT predictions of the proton charge FF $G_E^p(Q^2)$ 
\cite{Alarcon:2018irp} for several values of the proton radius, $r_E^p = (0.80, 0.82, 0.84, 0.86, 0.88, 0.90)$ fm
(the values are indicated on the panels). 
The bands show the theoretical uncertainty resulting from the effective description of high-mass states in the 
spectral function (see text).
\textit{Dark shaded bands (labeled ``Global Fit''):} 
$G_E^p(Q^2)$ determined from global fits of the elastic FF data with constrained
proton radius (see text) \cite{Ye:2017gyb}. The bands show the experimental and fit uncertainties.
Each panel's global fit was restricted to reproduce the indicated proton radius.}
}
\label{fig:dicheft}
\end{figure*}
\section{Method}
\subsection{Predictive theoretical framework}
DI$\chi$EFT is a method for calculating the nucleon FFs combining chiral effective field theory 
($\chi$EFT) --- a systematic description of strong interactions at distances $\mathcal{O}(M_\pi^{-1})$;
and dispersion analysis --- the use of complex analyticity for connecting the behavior of 
hadronic amplitudes in different kinematic regions. The method is described in detail in
Refs.~\cite{Alarcon:2017ivh,Alarcon:2017lhg,Alarcon:2018irp}; the essential elements 
used in the present calculation are summarized for reference in Appendix~\ref{app:parametrization}.
The FFs are represented as dispersive integrals over $t \equiv - Q^2$. The spectral functions 
on the two-pion cut at $t > 4 M_\pi^2$ are calculated using (i)~the elastic unitarity relation;
(ii)~$\pi N$ amplitudes computed in $\chi$EFT at leading order, next-to-leading order, and
partial next-to-next-to-leading order accuracy; (iii)~the timelike pion FF measured in $e^+e^-$ 
annihilation experiments. The approach includes $\pi\pi$ rescattering effects and the $\rho$ resonance
and allows one to calculate the two-pion spectral functions up to $t \approx 1$ GeV$^2$.
Higher-mass $t$-channel states are described by effective poles, whose strength is fixed by the
dispersive integrals for the nucleon charges, magnetic moments, and radii (sum rules) \cite{Alarcon:2018irp}.
The nucleon radii thus enter as explicit parameters in the DI$\chi$EFT predictions of the
spectral functions. Evaluating the finite-$Q^2$ dispersive integrals with these spectral functions, 
we obtain an analytic parametrization of the spacelike FFs in which the nucleon radii appear 
as explicit parameters [see Appendix~\ref{app:parametrization}, specifically Eqs.~(\ref{parametrization_proton}) 
and (\ref{parametrization_neutron})]; all other dynamical input is determined independently by the 
$\pi N$ scattering data and the pion timelike FF data. We emphasize that in this approach
the correlation between the radius and the finite-$Q^2$ behavior appears through the 
global analytic properties of the FF (dispersive representation, sum rules), not through a 
power series expansion at $Q^2 = 0$; the correlation therefore extends beyond the range of
convergence of the power series expansion (the connection with the power 
series expansion is discussed further in Sec.~\ref{subsec:low_q2}).

In the application here we take the proton charge radius as a free parameter, to be varied over a range 
covering the presently discussed values. The neutron charge radius, which enters indirectly through the separate 
sum rules for the isovector and isoscalar FFs, is fixed at its Particle Data Group 
(PDG) value \cite{Tanabashi:2018oca}; its influence on the proton FF is negligible in the 
$Q^2$ region considered here (see Appendix~\ref{app:parametrization}, specifically Fig.~\ref{fig:gep_param}).
In this way DI$\chi$EFT provides us with a family of theoretical predictions
of $G_E^p(Q^2)$, with each function respecting analyticity in $Q^2$ and corresponding to definite 
value of the proton charge radius. Figure~\ref{fig:dicheft} shows the predictions
for a set of radii $r_E^p = (0.80, 0.82, 0.84, 0.86, 0.88, 0.90)$ fm.

The dominant uncertainties in the DI$\chi$EFT predictions of $G_E^p (Q^2)$ (for a given $r_E^p$) 
arise from the the effective description of high-mass states in the isovector spectral function.
We have estimated them conservatively, by varying the position of the effective pole 
over a range $M_1^2 =$ (0.5--2)$\times M_1^2(\textrm{nom})$, 
where $M_1^2(\textrm{nom})$ is the nominal value determined in Ref.~\cite{Alarcon:2018irp}
[see Appendix~\ref{app:parametrization}, specifically Eq.~(\ref{spectral_isovector_highmass})].
The results are shown by the bands in Fig.~\ref{fig:dicheft}. One sees that the uncertainties
of the FF predictions are small at low $Q^2$ (where the dispersive integral is dominated by the two-pion cut
and constrained by the given value of the radius), but increase at larger $Q^2$ (where the dispersive 
integral becomes sensitive to high-mass states in the spectral function).

\subsection{Descriptive global fit}
In order to confront the DI$\chi$EFT predictions with the experimental proton FF we use the 
results of a descriptive global fit \cite{Ye:2017gyb}. It employs a bounded 
polynomial $z$-expansion \cite{Hill:2010yb} and determines $G_E^p$ and $G_M^p$ 
directly from the cross section and polarization data. Sum rules are imposed
to ensure the correct normalization at $Q^2=0$ and the asymptotic scaling behavior 
at large $Q^2$ \cite{Lee:2015jqa}. The treatment of uncertainties includes the 
covariance matrix of the fit itself, the systematic errors arising from the tension 
between data sets, and the uncertainty from two-photon exchange
corrections at high $Q^2$. In the original work of Ref.~\cite{Ye:2017gyb}
the proton radii in the fit were fixed at their presumed empirical values ($r_E^p$ = 0.879~fm from CODATA, 
$r_M^p$ = 0.851 fm from PDG); the radius constraints were then removed when evaluating the fit uncertainty. 
In the present study we have used the same fitting method to generate a family of fits with different 
values of the proton charge radius, covering the range 0.8--0.9 fm (see Fig.~\ref{fig:dicheft}).
The magnetic radius is kept fixed at its PDG value; the uncertainty resulting from this
simplification is negligible and covered by the quoted overall fit uncertainty.
In these fits we only include the uncertainties from the covariance matrix. One observes that in the global 
fits there is little correlation between the proton charge radius and the value of the FFs at finite $Q^2$ 
(the present data cover the range $\gtrsim$ 0.01 GeV$^2$), as expected for this descriptive approach.

\subsection{Method for radius extraction}
%
%
\begin{figure}[t]
\includegraphics[width=1.0\columnwidth]{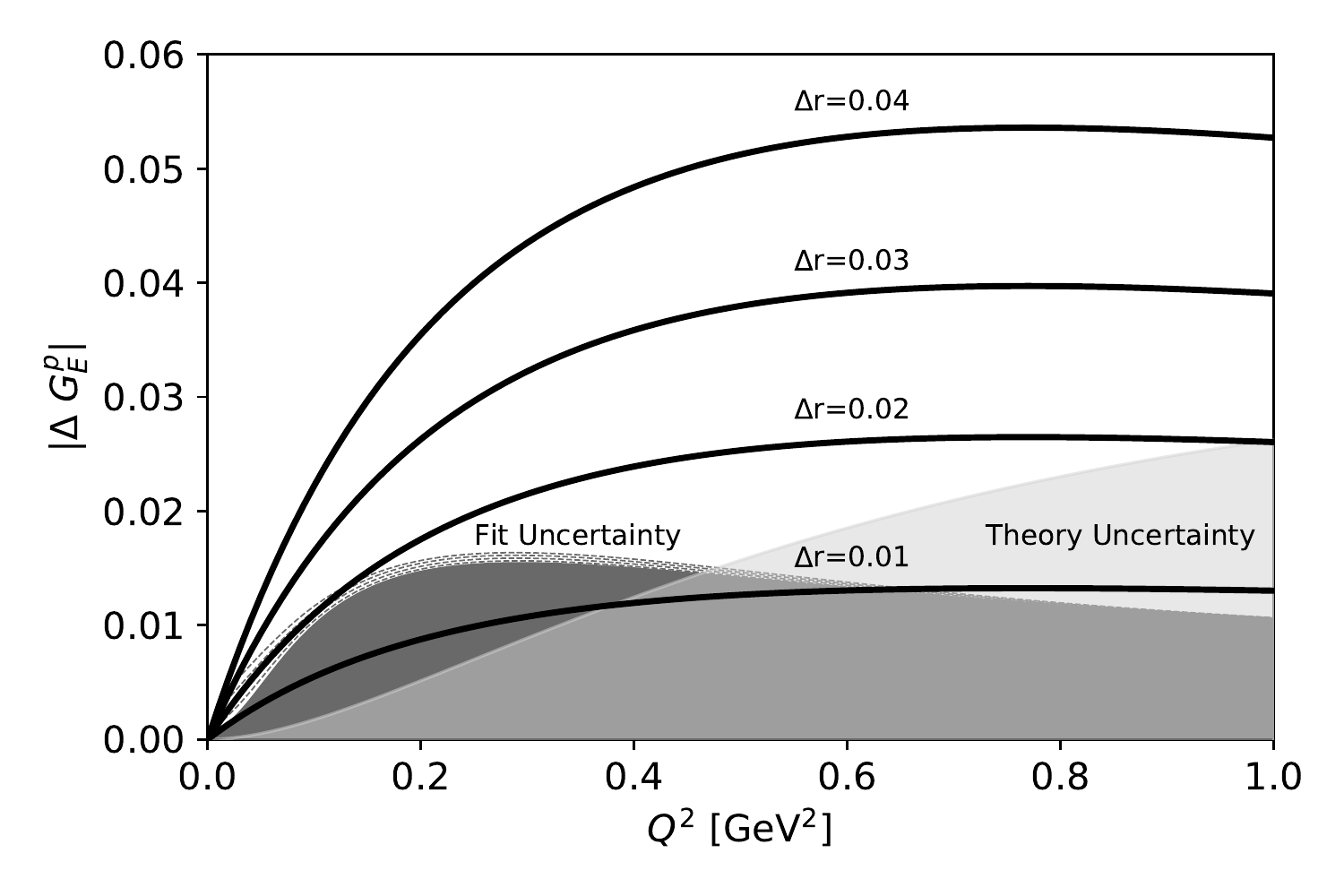}
\caption{\textit{Light shaded band (labeled ``Theory Uncertainty''):} Theoretical uncertainty of the 
DI$\chi$EFT predictions of $G_E^p(Q^2)$ for fixed $r_E^p$ (here $r_E^p =$ 0.84 fm;
the uncertainty does not depend on the specific value of $r_E^p$; cf.\ Fig.~\ref{fig:dicheft}) 
\cite{Alarcon:2018irp}. 
\textit{Dark shaded band and dashed area above (labeled ``Fit Uncertainty''):} Experimental uncertainty 
of the global FF fit \cite{Ye:2017gyb} (68\% confidence level, cf.\ Fig.~\ref{fig:dicheft}). 
The dark shaded band shows the uncertainty of the fits with a fixed proton radius, 
as shown in the individual panels of Fig.~\ref{fig:dicheft}. 
The dashed area above shows in addition the variation of the global fit under changes of the radius.
\textit{Solid lines:} Variation of the DI$\chi$EFT predictions of $G_E^p(Q^2)$ 
under changes of $r_E^p$ by $\Delta r = \pm$0.01~fm, $\pm$0.02~fm, etc.\ (the variation is computed 
at $r_E^p =$ 0.84 fm). The variation quantifies the sensitivity of the theoretical FF predictions
to the proton radius parameter (cf.\ Fig.~\ref{fig:dicheft}).}
\label{fig:delta_ge}
\end{figure}
Figure~\ref{fig:dicheft} now allows us to compare the DI$\chi$EFT predictions with the global fits to
the FF data for several assumed values of the proton radius. We observe: (a)~Whereas the global fits show
little correlation between the radius and the finite-$Q^2$ FF, the DI$\chi$EFT predictions show
a strong correlation, as a consequence of the analytic properties. (b)~There is a clearly preferred value of the radius
at $\sim$0.84 fm, for which there is best agreement of the DI$\chi$EFT predictions with the global fit
to the FF data. (c)~At the radius of best agreement, the DI$\chi$EFT predictions 
provide a description of the data with the same accuracy as the global fit up to $Q^2 \sim 1$ GeV$^2$
(actually up to even larger values $Q^2 \sim 2$ GeV$^2$, which are not included in Fig.~\ref{fig:dicheft}).

The observations suggest a simple method for extracting the proton radius from the FF data:
Compare the DI$\chi$EFT predictions with the global fit in the region where both descriptions 
are valid, and determine the radius by best agreement. The method combines the advantages
of the descriptive fit (reliable uncertainty estimates in the region where there are data) 
and the predictive theory (correlation of the radius and the finite-$Q^2$ behavior 
through analyticity). 

The method can be optimized by choosing the ``best'' $Q^2$ region for the comparison.
Figure~\ref{fig:delta_ge} shows the theoretical uncertainty of the DI$\chi$EFT FF prediction
for a fixed proton radius, and the experimental uncertainties of the FF obtained from the global fit.
It also shows the variation of the DI$\chi$EFT FF predictions under a certain change of the proton 
radius (here, $\Delta r = \pm$0.01~fm, $\pm$0.02~fm, etc.), which quantifies the sensitivity 
of the FF to the proton radius as a function of $Q^2$. 
One observes: (a)~At ``low'' $Q^2$ ($\sim$0.1 GeV$^2$) the theoretical uncertainty of the 
DI$\chi$EFT FF is much smaller than the experimental uncertainty of the global fit, which 
is mainly due to normalization errors (inconsistencies between data sets).
The variation of the FF under the change $\Delta r = \pm 0.01$ is larger than the theoretical 
uncertainty but smaller than the fit uncertainty. (b)~At ``high'' $Q^2$ ($\sim$1 GeV$^2$) the 
theoretical uncertainty is larger than the fit uncertainty. The variation of the FF under 
$\Delta r = \pm 0.01$ is smaller than the theoretical uncertainty and comparable to the
fit uncertainty. Based on Fig.~\ref{fig:delta_ge} we choose the upper limit of the $Q^2$-region 
for radius extraction as $Q^2_{\rm max} \sim 0.5$ GeV$^2$; with this upper limit the theoretical error
remains smaller than, or at most comparable to, the fit error. 
The lower limit we choose as $Q^2_{\rm min} \sim$ 0.01 GeV$^2$; 
this represents the lowest value for which FF data are presently available, and the results 
are not sensitive to this choice. (The performance of our method with smaller values of 
$Q^2_{\rm max}$ and the relation to low-$Q^2$ fits using a power series expansion are 
discussed in Sec.~\ref{subsec:low_q2}.)

\begin{figure}[t]
\includegraphics[width=\columnwidth]{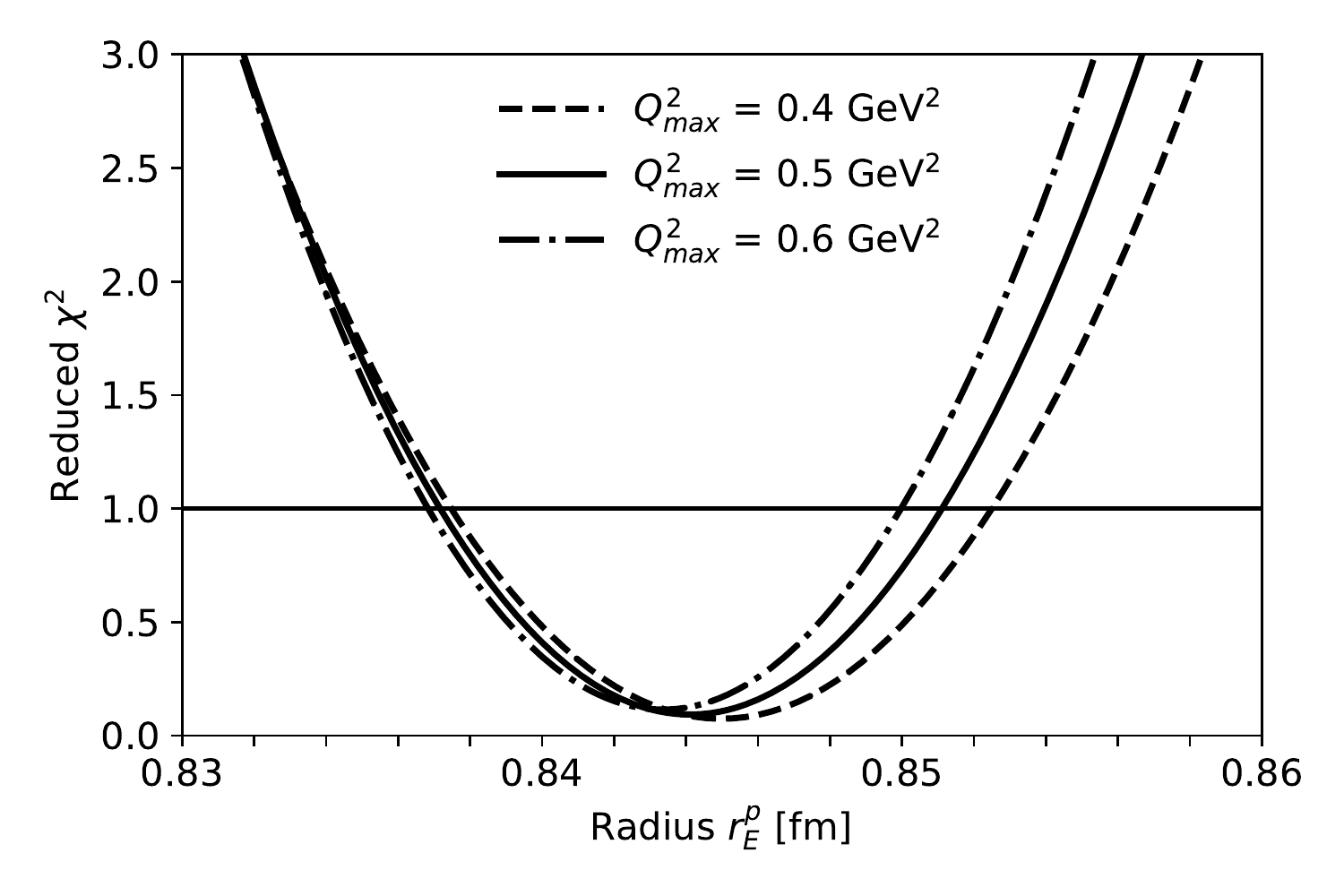}
\caption{Reduced $\chi^2$, Eq.~(\ref{chi2}), as a function of the
proton radius $r_E^p$. Shown are the results corresponding to three choices 
of the upper limit $Q^2_{\rm max}$.}
\label{fig:chi2}
\end{figure}

To quantify the agreement of the theoretical model with the global fit and extract the radius,
we use a figure of merit in the form of a reduced $\chi^2$,
\begin{eqnarray}
\chi^2(r_E^p) &\equiv& N^{-1}
\sum_{\textrm{bins $i$}} \frac{(\textrm{thy}_i - \textrm{fit}_i)^2}{(\Delta\textrm{thy}_i)^2 + 
(\Delta\textrm{fit}_i)^2}
\nonumber
\\[1ex]
&& \left\{ \textrm{thy}_i \equiv G_E^p(Q^2_i) \; [\textrm{DI$\chi$EFT, given $r_E^p$}], \right.
\nonumber
\\[0ex]
&& \left. \;\; \textrm{fit}_i \equiv G_E^p(Q^2_i) \; [\textrm{global fit, given $r_E^p$}] \right\} .
\label{chi2}
\end{eqnarray}
The sum runs over $N$ bins in $Q^2$ covering the range $(Q^2_{\rm min}, Q^2_{\rm max})$; we use
$N =$ 50; the results are not sensitive to the binning. $\textrm{thy}_i$ denotes the 
DI$\chi$EFT prediction of $G_E^p(Q_i^2)$ in the $i$'th $Q^2$ bin for the given $r_E^p$; 
$\textrm{fit}_i$ denotes the global fit result for $G_E^p(Q_i^2)$ in the same bin.
The theoretical and fit uncertainties, $\Delta\textrm{thy}_i$ and $\Delta\textrm{fit}_i$,
are added in quadrature. Figure~\ref{fig:chi2} shows the reduced $\chi^2$ as a function of $r_E^p$. 
One observes: (a) The dependence is approximately quadratic, indicating 
a natural best agreement without tension. (b)~If the uncertainty of $r_E^p$ is 
defined by the criterion $\Delta\chi^2 \leq 1$, the results obtained with 
$Q^2_{\rm max} =$ 0.4, 0.5, and 0.6 GeV$^2$ are consistent within uncertainties, 
affirming our choice of the optimal $Q^2_{\rm max}$.
%
%
\section{Results}
Using the above method and $Q^2$-range, we have extracted the proton charge radius and 
its uncertainty, obtaining a value $r_E^p$ = 0.844(7) fm. The uncertainty
estimate is based on the combined theoretical and global fit uncertainties entering 
in our figure of merit Eq.~(\ref{chi2}) and corresponds to a confidence interval with
$\Delta \chi^2 \leq 1$. The extracted radius is consistent with the high-precision muonic
hydrogen results and clearly disfavors the CODATA result. Our result therefore
suggests that the finite-$Q^2$ electron scattering data agree well with the 
muonic hydrogen results, and that the disagreement is rather between the
electronic and muonic hydrogen results. We note that some recent measurements 
with electronic hydrogen have yielded a value consistent with the muonic 
result \cite{Beyer79}, while others agree with the current CODATA 
value~\cite{Fleurbaey:2018fih}.
\section{Discussion}
\subsection{Possible improvements}
The proton radius extraction reported here could be improved in several aspects:
(a) by reducing the theoretical uncertainty of the DI$\chi$EFT FF predictions through an improved 
description of the high-mass spectral functions ($t >$ 1 GeV$^2$), which would be possible with 
a more flexible parametrization and further theoretical constraints; (b)~by reducing the experimental 
uncertainties of the FF data, especially in the region $Q^2 \lesssim$ 0.2 GeV$^2$, which would 
allow us to limit the theory-experiment comparison to a smaller $Q^2$-interval
($Q^2_{\rm max} \sim$ 0.2 GeV$^2$), where the theoretical uncertainties are smaller.

In our analyticity-based framework the main impact on the proton radius comes from FF data 
at moderate $Q^2$ ($\sim$ 0.1--0.5 GeV$^2$) rather than at the lowest available $Q^2$. 
The forthcoming FF data at very low $Q^2$ from the Jefferson Lab PRad 
experiment \cite{Gasparian:2014rna} (down to a few $\times 10^{-4}$ GeV$^2$) can complement the 
results of our study by reducing the normalization errors of the data (cf.\ the discussion below)
and enabling an independent radius extraction using traditional extrapolation methods. 
They can also validate our theoretical framework, e.g.\ by extracting higher FF derivatives, 
which enable sensitive tests of the DI$\chi$EFT spectral functions \cite{Alarcon:2017lhg}.

\subsection{Relation to low-$Q^2$ FF fits}
\label{subsec:low_q2}
%
%
\begin{figure}[t]
\includegraphics[width=\columnwidth]{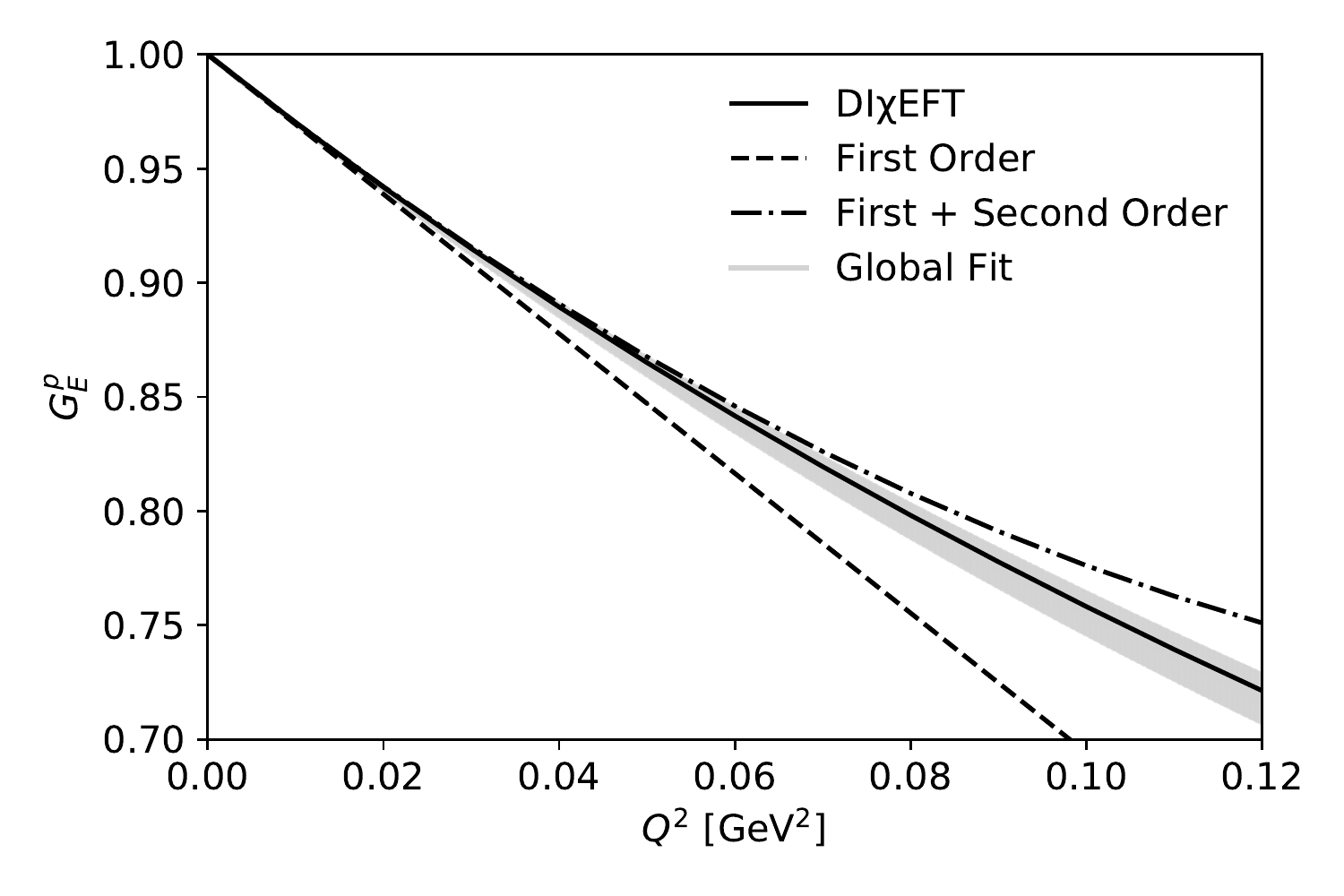}
\caption{Low-$Q^2$ behavior of the DI$\chi$EFT parametrization of $G_E^p(Q^2)$
and its power expansion. \textit{Solid line:} 
DI$\chi$EFT parametrization ($r_E^p = 0.844$ fm). \textit{Dashed line:} First-order term in $Q^2$.
\textit{Dashed-dotted line:} Sum of first- and second-order terms. \textit{Shaded band:}
Uncertainty of the global fit with the given fixed proton radius.}
\label{fig:moments}
\end{figure}
In Ref.~\cite{Horbatsch:2016ilr} the proton radius was extracted from fits to the low-$Q^2$
FF data ($Q^2_{\rm max} \lesssim 0.02$ GeV$^2$) with a truncated power series in $Q^2$, 
in which the coefficient of the first-order term in $Q^2$ (the proton radius) was determined 
by the data, and the coefficients of the higher-order terms $Q^4, Q^6,$ etc.\ (the higher moments) 
were calculated in standard $\chi$EFT and supplied as a theoretical input.
It is worth explaining how our dispersive method relates to that approach when the fit 
is restricted to the low-$Q^2$ region. Figure~\ref{fig:moments} shows the DI$\chi$EFT FF
parametrization (with the radius determined by our above best fit), as well as its first-order 
and second-order expansion in $Q^2$; the coefficients are given by the derivatives of 
the FF, evaluated using the dispersive integral with the DI$\chi$EFT spectral functions. 
One sees that the full DI$\chi$EFT FF is well approximated by the second-order expansion, with the 
second-order term giving a correction of 1\% (5\%) at $Q^2 =$ 0.03 GeV$^2$ (0.06 GeV$^2$). 
The second derivative of the DI$\chi$EFT FF changes only by a few percent when the radius is varied 
within the range $r^p_E$ = (0.80, 0.90) fm, so that the coefficient of the second-order term may be regarded
as a fixed theoretical input. Our method therefore effectively reduces to that of Ref.~\cite{Horbatsch:2016ilr}
in the region where the second-order approximation is valid ($Q^2_{\rm max} < 0.05$ GeV$^2$).
The advantage of our method is that it is not limited by the power series expansion and allows us to 
include FF data at significantly higher $Q^2$ into the fit, making the radius extraction 
more robust. This can be seen in the trend shown in Fig.~\ref{fig:chi2}: increasing the value of
$Q^2_{\rm max}$ decreases the uncertainty in the radius, as the fit is constrained by more data,
while at the same time the theoretical uncertainties are still under control. For reference we note that, 
if we restricted our fit to $Q^2_{\rm max} = 0.1$ GeV$^2$, we would obtain a radius $r^p_E$ = 0.849(10) fm,
which agrees well with the result of Ref.~\cite{Horbatsch:2016ilr} in the central value and the uncertainty.

We also point out that the second derivative (or moment) of the DI$\chi$EFT FF is significantly larger 
than the standard $\chi$EFT results used in Ref.~\cite{Horbatsch:2016ilr}, 
because the $\pi\pi$ rescattering effects included in the DI$\chi$EFT calculation
increase the spectral function on the two-pion cut in the near-threshold region; 
see the discussion in Ref.~\cite{Alarcon:2017lhg}.
This shows that in the approach of Ref.~\cite{Horbatsch:2016ilr} the theoretical uncertainty
would deteriorate quickly if it were applied at $Q^2$ values where the second-order 
(and higher-order) terms become sizable. We note that the $\pi\pi$ rescattering effects
are noticeable even in the higher FF derivatives ($n \geq 3$); nevertheless, for these
quantities the DI$\chi$EFT results agree with the standard $\chi$EFT predictions within 
uncertainties; see Ref.~\cite{Alarcon:2017lhg} for details.
Explicit expressions for the FF derivatives in standard $\chi$EFT can be found in Ref.~\cite{Peset:2014jxa}.

Some comments are in order regarding the experimental uncertainties in FF fits at low 
$Q^2$ ($\sim 0.01$ GeV$^2$). The dominant uncertainty in this case results from the 
normalization errors of the FF data. In the fits the normalization of
different data sets is adjusted, using the given fit function or theoretical model, 
and requiring that $G_E^p(0) = 1$. Excellent fits can be achieved after this rescaling
of the data sets, and one may be tempted to conclude that the radius could be extracted 
with great precision from them. However, such reasoning would be circular: The rescaling
of the data sets depends on the fit function, and at low $Q^2$ the behavior of that 
function is governed by the radius, so that the procedure essentially forces the 
rescaled data to reproduce the assumed radius, resulting in a loss of sensitivity to the radius.
This effect can be seen in the global fit uncertainties shown in Fig.~\ref{fig:delta_ge}.
The dark shaded band shows the uncertainty of the global fit for a certain fixed value of 
the radius, i.e., constraining the first-order term in the fit function. 
This uncertainty vanishes rapidly in the limit $Q^2 \rightarrow 0$, reflecting the trivial 
fact that the constrained fit with an assumed radius reproduces that value of the radius.
The dashed area above the dark shaded band shows the variation in the global fit result under
a change of the radius by $\Delta r = \pm (0.01, 0.02, 0.03, 0.04)$ fm. The combined area 
represents the actual experimental uncertainty in the radius extraction in low-$Q^2$ fits.
It is much larger than the fixed-radius uncertainty at low $Q^2$ and vanishes much more slowly
in the $Q^2 \rightarrow 0$ limit. Altogether, this points to a principal limitation
of radius extraction from fits to low-$Q^2$ FF data. Our DI$\chi$EFT method overcomes this
limitation by recruiting higher-$Q^2$ data for the radius extraction 
(up to $Q^2_{\rm max} \sim 0.5$ GeV$^2$), whose impact is only minimally 
affected by normalization errors.

\subsection{Relation to empirical dispersive fits}
The proton radius was extracted previously from dispersive FF fits in which the two-pion spectral 
functions were constructed by analytic continuation of empirical $\pi N$ amplitudes \cite{Belushkin:2006qa,Lorenz:2012tm}. 
In these approaches the two-pion spectral functions are completely determined before they are
placed in the dispersive integrals and used to evaluate FFs and radii. 
Our method is different in that the two-pion spectral functions are computed in DI$\chi$EFT and 
contain an unknown low-energy constant (related to the nucleon radii, cf.\ Appendix~\ref{app:parametrization}), 
which can vary and adjust the strength of the spectral functions in the $\rho$ meson peak and 
above \cite{Alarcon:2018irp}. This increases the flexibility of the FF description and enables a
more robust radius extraction. We point out that the DI$\chi$EFT spectral functions at partial N2LO accuracy, 
evaluated with a realistic range of proton radii, agree very well with those of the Roy-Steiner 
analysis of Ref.~\cite{Hoferichter:2016duk}, but differ significantly from those of 
Refs.~\cite{Belushkin:2006qa,Lorenz:2012tm} in the $\rho$ meson mass region; 
see Ref.~\cite{Alarcon:2018irp} for a detailed comparison. Even so, the empirical dispersive 
fits have consistently obtained proton radii $\sim$~0.84 fm \cite{Belushkin:2006qa,Lorenz:2012tm}, 
in agreement with our result.
\appendix
\section{Form factor parametrization}
\label{app:parametrization}
In this appendix we summarize the DI$\chi$EFT calculation of the nucleon FFs and describe
the parametrization used in the radius extraction in the present work. Further information about the
method and other applications can be found in Refs.~\cite{Alarcon:2017ivh,Alarcon:2017lhg,Alarcon:2018irp}.

The nucleon electric FFs are separated into isovector and isoscalar components, 
\beq
G_E^{p, n} = \pm G_E^V + G_E^S , \hspace{2em}
G_E^{V, S} \equiv {\textstyle\frac{1}{2}}(G_E^p \mp G_E^n),
\eeq 
and represented as dispersive integrals over $t \equiv -Q^2$, 
\beq
G^{V, S}_E (t) \; = \; \frac{1}{\pi} \int_{t_{\rm thr}}^\infty dt' \; 
\frac{\text{Im} \, G^{V, S}_E(t')}{t' - t - i 0} .
\label{dispersive_representation}
\eeq
The integrands involve the imaginary parts of the FFs on the cuts at $t' > t_{\rm thr} > 0$,
which are known as the spectral functions.
In the isovector component $t_{\rm thr} = 4 M_\pi^2$, and the spectral function is organized as
\beq
\text{Im} \, G^{V}_E(t') \; = \; 
\text{Im} \, G^{V}_E(t')[\pi\pi] \; + \; 
\text{Im} \, G^{V}_E(t')[\textrm{high-mass}] .
\label{isovector_spectral_low_high}
\eeq
The first term accounts for the contribution of the two-pion cut from $t_{\rm thr}$
up to $t_{\rm max} \sim$ 1 GeV$^2$ and is
calculated theoretically using the elastic unitarity relation in the $\pi\pi$ channel, 
the $\pi N$ amplitudes computed in $\chi$EFT, and the empirical timelike pion FF; 
the explicit expressions are given in 
Refs.~\cite{Alarcon:2017lhg,Alarcon:2018irp}. The couplings entering in the $\chi$EFT amplitudes
at LO and NLO accuracy are determined by pion-nucleon scattering data. 
At partial N2LO accuracy the $\chi$EFT result involves one unknown low-energy constant, $\lambda$, 
which represents a free parameter; schematically
\beq
\text{Im} \, G^{V}_E[\pi\pi] \; = \; \textrm{[LO]} + \textrm{[NLO]} + \lambda \, \textrm{[N2LO]} .
\label{spectral_isovector_pipi}
\eeq
The second term in Eq.~(\ref{isovector_spectral_low_high}) accounts for the high-mass states in the 
spectral function above $t_{\rm max}$ and is parametrized by an effective pole,
\beq
\text{Im} \, G^V_E(t')[\textrm{high-mass}] \; = \; \pi a^{(1)}_E \, \delta(t' - M_{1}^2) ,
\label{spectral_isovector_highmass}
\eeq
where the pole position $M_1^2 = 2.1\, \textrm{GeV}^2$ is inferred from the $e^+e^-$ annihilation data
and the pole strength represents a free parameter; the justification for this approximation and
its accuracy are discussed in Ref.~\cite{Alarcon:2018irp}. The values of the parameters $\lambda$ 
and $a_E^{(1)}$ in Eqs.~(\ref{spectral_isovector_pipi}) and (\ref{spectral_isovector_highmass})
are fixed by the sum rules for the nucleon isovector electric charge and radius,
\be
\frac{1}{\pi}\int_{t_{\rm thr}}^\infty dt' \, \frac{\text{Im} \, G_E^V(t')}{t'} 
&=& Q_E^V = {\textstyle \frac{1}{2}} , 
\label{sumrule_charge} \\
 \frac{6}{\pi}\int_{t_{\rm thr}}^\infty dt' \, \frac{\text{Im} \, G_E^V(t')}{t'^2} 
&=& (r_E^V)^2
\equiv {\textstyle \frac{1}{2}} [ (r_E^p)^2 - (r_E^n)^2 ] .
\hspace{2em}
\label{sumrule_radius}
\ee
Since the value of the isovector charge is known, this leaves the isovector radius as the only 
free parameter of the isovector spectral function Eq.~(\ref{isovector_spectral_low_high}).
Note that the relation between the original parameters and the charge and radius implied by
Eqs.~(\ref{sumrule_charge}) and (\ref{sumrule_radius}) is linear,
\beq
\{\lambda, a_E^{(1)} \} \; \stackrel{\textrm{linear}}{\longleftrightarrow} \; \{ Q_E^V, (r_E^V)^2 \}.
\label{linear}
\eeq
Expressing the parameters in terms of the radius, substituting the spectral function 
in Eq.~(\ref{dispersive_representation}), and performing the dispersive integral, 
we obtain a parametrization of the spacelike isovector FF ($t < 0$) in the form
\beq
G_E^V(t) \; = \; A_E^V(t) \, + \, (r_E^V)^2 \, B_E^V(t) .
\label{parametrization_isovector}
\eeq
The functions $A_E^V(t)$ and $B_E^V(t)$ represent, respectively, the dispersive integrals over the parts of the 
spectral function that are independent of $(r_E^V)^2$, and proportional to 
$(r_E^V)^2$. The functions have the same analytic structure as the full FF
and embody the full complex $t$ dependence in the spacelike region, as dictated by the dispersive representation.
Note that the linear decomposition in Eq.~(\ref{parametrization_isovector}) results from the linear relation 
of $(r_E^V)^2$ to the original theoretical parameters; it does not imply an expansion in $t$
or other approximation to the complex $t$-dependence of the FF.

In the isoscalar component of Eq.~(\ref{dispersive_representation}) $t_{\rm thr} = 9 M_\pi^2$,
and the spectral function is parametrized as the sum of two effective poles,
\beq
\text{Im} \, G^S_E(t') \, = \, \sum_{r = \omega, \phi}
\pi a^r_E \delta(t' - M_r^2) .
\label{spectral_isoscalar}
\eeq
The first pole is at the $\omega$ resonance mass and accounts for the $3\pi$ strength; the second pole is 
at the $\phi$ mass and effectively accounts for the $\phi$ resonance and other hadronic contributions at 
higher masses. The parameters $a_E^\omega$ and $a_E^\phi$ are fixed by the sum rules for the nucleon's isoscalar 
electric charge and radius,
\be
\frac{1}{\pi}\int_{t_{\rm thr}}^\infty dt' \, \frac{\text{Im} \, G_E^S(t')}{t'} 
&=& Q_E^S = {\textstyle \frac{1}{2}} , 
\label{sumrule_charge_isoscalar} \\
 \frac{6}{\pi}\int_{t_{\rm thr}}^\infty dt' \, \frac{\text{Im} \, G_E^S(t')}{t'^2} 
&=& (r_E^S)^2
\equiv {\textstyle \frac{1}{2}} [(r_E^p)^2 + (r_E^n)^2 ] ,
\hspace{3em}
\label{sumrule_radius_isoscalar}
\ee
which imply a linear relation similar to Eq.~(\ref{linear}). Altogether we obtain a representation of the 
spacelike isoscalar FF analogous to the isovector case, Eq.~(\ref{parametrization_isovector}),
\beq
G_E^S(t) \; = \; A_E^S(t) \, + \, (r_E^S)^2 \, B_E^S(t) .
\label{parametrization_isoscalar}
\eeq

Combining the isovector and isoscalar parametrizations, Eqs.~(\ref{parametrization_isovector})
and Eqs.~(\ref{parametrization_isoscalar}), we obtain a parametrization of the proton an neutron
FFs in terms of the isovector and isoscalar radii,
\be
G_E^{p, n}(t) \; &=& \; \pm A_E^V(t) \, + \, A_E^S(t) 
\nonumber \\
&& \pm (r_E^V)^2 \, B_E^V(t) \, + \, (r_E^S)^2 \, B_E^S(t) .
\ee
It may also be expressed directly in terms of the individual nucleon radii,
\be
G_E^p(t) \, &=& \, A_E^p (t) \, + \, (r_E^p)^2 \, B_E(t) \, + \, (r_E^n)^2 \, \bar{B}_E(t) , \hspace{1em}
\hspace{1em}
\label{parametrization_proton}
\\[0ex]
G_E^n(t) \, &=& \, A_E^n (t) \, + \, (r_E^n)^2 \, B_E(t) \, + \, (r_E^p)^2 \, \bar{B}_E(t) , 
\hspace{1em}
\label{parametrization_neutron}
\ee
where the functions are
\be
A_E^{p, n} &\equiv& \pm A_E^V + A_E^S ,
\\[0ex]
B_E &\equiv& {\textstyle\frac{1}{2}} (B_E^V + B_E^S) ,
\\[0ex]
\bar{B}_E &\equiv& {\textstyle\frac{1}{2}} (-B_E^V + B_E^S) .
\ee
The functions $A^{p, n}(t)$ describe the radius-independent part of the nucleon FFs 
(in the context of the dispersive parametrization defined above);
the functions $B_E(t)$ and $\bar{B}_E(t)$ describe, respectively, the parts proportional to the
charge radii of the ``same'' and the ``other'' nucleon, under the constraints of isospin symmetry.
Several properties of the functions follow immediately from their definitions:
\beq
\left.
\begin{array}{rclrcl}
A_E^{p, n}(0) &=& Q_E^{p, n},  \hspace{2em} & dA_E^{p, n}/dt(0) &=& 0, 
\\[1ex]
B_E(0) &=& 0, & 6 \; dB_E/dt(0) &=& 1,
\\[1ex]
\bar{B}_E(0) &=& 0, & d\bar{B}_E/dt(0) &=& 0,
\end{array} \hspace{1em}
\right\}
\label{A_B_C_properties}
\eeq
where $Q_E^{p, n} = \{ 1, 0\}$ is the nucleon electric charge. In particular, these conditions
ensure that the nucleon FFs described by Eqs.~(\ref{parametrization_proton}) and (\ref{parametrization_neutron}) 
have the correct values at zero momentum transfer, $G_E^{p, n}(0) = Q_E^{p, n}$, and that the first derivatives 
of the nucleon FFs are controlled exclusively by the radii of the ``same'' nucleon,
\be
6 \, d G_E^{p, n}/dt (0) &=& 6 \; d B_E/dt (0) \, (r_E^{p, n})^2 
\; = \; 6 \, (r_E^{p, n})^2 . \hspace{3em}
\ee

%
%
\begin{figure}[t]
\includegraphics[width=\columnwidth]{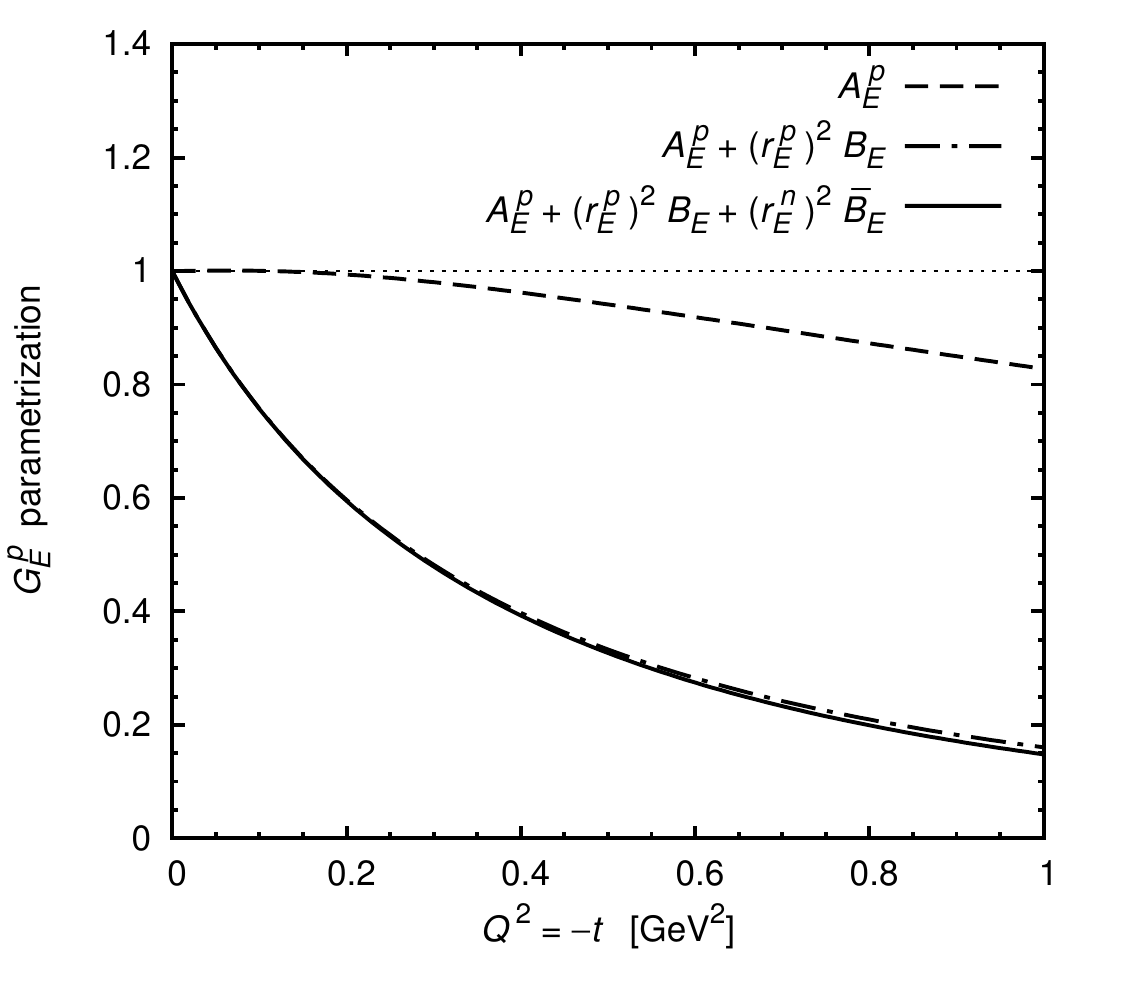}
\caption{Proton FF parametrization Eq.~(\ref{parametrization_proton}) as function of $Q^2 = -t$. 
Dashed line: $A_E^p(t)$. Dashed-dotted line: 
$A_E^p(t) + (r_E^p)^2 \, B_E(t)$. Solid line:
$A_E^p(t) +  (r_E^p)^2 \, B_E(t) + (r_E^n)^2 \, \bar{B}_E(t)$ (full result).}
\label{fig:gep_param}
\end{figure}
Figure~\ref{fig:gep_param} shows the contributions of the individual terms in the proton FF parametrization
Eq.~(\ref{parametrization_proton}) as functions of $Q^2 = -t$. One observes: (a)~The radius-independent 
term $A_E^p$ starts with value 1 and derivative 0 at $Q^2 = 0$, cf.\ Eq.~(\ref{A_B_C_properties});
its value remains close to 1 for all $Q^2 < 1$ GeV$^2$. (b)~The proton-radius-dependent term $ (r_E^p)^2 \, B_E$ 
starts with value 0 at $Q^2 = 0$ and accounts for the derivative of $G_E^p$ at $Q^2 = 0$; it causes most 
of the $Q^2$-dependence of the FF (i.e., the deviation of $G_E^p$ from 1) at $Q^2 < 1$ GeV$^2$. 
(c)~The neutron-radius-dependent term $ (r_E^n)^2 \, \bar{B}_E$ starts with value 0 and 
derivative 0 at $Q^2 = 0$; its contribution to $G_E$ remains $<0.01$ for $Q^2 < 1$ GeV$^2$. 
Altogether, the sum $A_E^p +  (r_E^p)^2 \, B_E$ practically accounts for the entire value 
of the FF and its $Q^2$-dependence all at $Q^2 < 1$ GeV$^2$; the $ (r_E^n)^2 \, \bar{B}_E$ term
represents a percent-level correction. This makes the parametrization Eq.~(\ref{parametrization_proton})
particularly useful for extracting the proton radius from the FF data. The parametrization of the
neutron FF Eq.~(\ref{parametrization_neutron}) follows the same pattern, with the role of the
radii reversed.

The numerical evaluation of the FFs with the radius-dependent dispersive FF parametrizations, 
Eqs.~(\ref{parametrization_proton}) and (\ref{parametrization_proton}), is extremely simple.
The functions $A_E^{p, n}(t)$, $B_E(t)$ and $\bar{B}_E(t)$ can be pre-computed and tabulated as
functions of $Q^2 = -t$. The FFs are then generated by multiplying these functions with the 
radii and combining the terms. This method was used to produce the plots of Fig.~\ref{fig:dicheft} 
and perform the radius extraction summarized in Fig.~\ref{fig:chi2}. The parametrization can be
used also in other studies of low-$Q^2$ proton and neutron FFs. Tables of the 
functions are are available upon request.
\begin{acknowledgments}
This material is based upon work supported by the U.S.~Department of Energy, 
Office of Science, Office of Nuclear Physics under contracts DE-AC05-06OR23177
and DE-AC02-06CH11357.
J.M.A.\ acknowledges support from the Community of Madrid through the
Programa de atracci\'on de talento investigador 2017 (Modalidad 1),
the Spanish MECD grants FPA2016-77313-P, FPA2016-75654-C2-2-P and the
group IPARCOS.
\end{acknowledgments}
\bibliography{elastic}

\begin{thebibliography}{33}%
\makeatletter
\providecommand \@ifxundefined [1]{%
 \@ifx{#1\undefined}
}%
\providecommand \@ifnum [1]{%
 \ifnum #1\expandafter \@firstoftwo
 \else \expandafter \@secondoftwo
 \fi
}%
\providecommand \@ifx [1]{%
 \ifx #1\expandafter \@firstoftwo
 \else \expandafter \@secondoftwo
 \fi
}%
\providecommand \natexlab [1]{#1}%
\providecommand \enquote  [1]{``#1''}%
\providecommand \bibnamefont  [1]{#1}%
\providecommand \bibfnamefont [1]{#1}%
\providecommand \citenamefont [1]{#1}%
\providecommand \href@noop [0]{\@secondoftwo}%
\providecommand \href [0]{\begingroup \@sanitize@url \@href}%
\providecommand \@href[1]{\@@startlink{#1}\@@href}%
\providecommand \@@href[1]{\endgroup#1\@@endlink}%
\providecommand \@sanitize@url [0]{\catcode `\\12\catcode `\$12\catcode
  `\&12\catcode `\#12\catcode `\^12\catcode `\_12\catcode `\%12\relax}%
\providecommand \@@startlink[1]{}%
\providecommand \@@endlink[0]{}%
\providecommand \url  [0]{\begingroup\@sanitize@url \@url }%
\providecommand \@url [1]{\endgroup\@href {#1}{\urlprefix }}%
\providecommand \urlprefix  [0]{URL }%
\providecommand \Eprint [0]{\href }%
\providecommand \doibase [0]{http://dx.doi.org/}%
\providecommand \selectlanguage [0]{\@gobble}%
\providecommand \bibinfo  [0]{\@secondoftwo}%
\providecommand \bibfield  [0]{\@secondoftwo}%
\providecommand \translation [1]{[#1]}%
\providecommand \BibitemOpen [0]{}%
\providecommand \bibitemStop [0]{}%
\providecommand \bibitemNoStop [0]{.\EOS\space}%
\providecommand \EOS [0]{\spacefactor3000\relax}%
\providecommand \BibitemShut  [1]{\csname bibitem#1\endcsname}%
\let\auto@bib@innerbib\@empty
\bibitem [{\citenamefont {Miller}(2019)}]{Miller:2018ybm}%
  \BibitemOpen
  \bibfield  {author} {\bibinfo {author} {\bibfnamefont {G.~A.}\ \bibnamefont
  {Miller}},\ }\href {\doibase 10.1103/PhysRevC.99.035202} {\bibfield
  {journal} {\bibinfo  {journal} {Phys. Rev.}\ }\textbf {\bibinfo {volume}
  {C99}},\ \bibinfo {pages} {035202} (\bibinfo {year} {2019})},\ \Eprint
  {http://arxiv.org/abs/1812.02714} {arXiv:1812.02714 [nucl-th]} \BibitemShut
  {NoStop}%
\bibitem [{\citenamefont {Punjabi}\ \emph {et~al.}(2015)\citenamefont
  {Punjabi}, \citenamefont {Perdrisat}, \citenamefont {Jones}, \citenamefont
  {Brash},\ and\ \citenamefont {Carlson}}]{Punjabi:2015bba}%
  \BibitemOpen
  \bibfield  {author} {\bibinfo {author} {\bibfnamefont {V.}~\bibnamefont
  {Punjabi}}, \bibinfo {author} {\bibfnamefont {C.~F.}\ \bibnamefont
  {Perdrisat}}, \bibinfo {author} {\bibfnamefont {M.~K.}\ \bibnamefont
  {Jones}}, \bibinfo {author} {\bibfnamefont {E.~J.}\ \bibnamefont {Brash}}, \
  and\ \bibinfo {author} {\bibfnamefont {C.~E.}\ \bibnamefont {Carlson}},\
  }\href {\doibase 10.1140/epja/i2015-15079-x} {\bibfield  {journal} {\bibinfo
  {journal} {Eur. Phys. J.}\ }\textbf {\bibinfo {volume} {A51}},\ \bibinfo
  {pages} {79} (\bibinfo {year} {2015})},\ \Eprint
  {http://arxiv.org/abs/1503.01452} {arXiv:1503.01452 [nucl-ex]} \BibitemShut
  {NoStop}%
\bibitem [{\citenamefont {Pacetti}\ \emph {et~al.}(2015)\citenamefont
  {Pacetti}, \citenamefont {Baldini~Ferroli},\ and\ \citenamefont
  {Tomasi-Gustafsson}}]{Pacetti:2015iqa}%
  \BibitemOpen
  \bibfield  {author} {\bibinfo {author} {\bibfnamefont {S.}~\bibnamefont
  {Pacetti}}, \bibinfo {author} {\bibfnamefont {R.}~\bibnamefont
  {Baldini~Ferroli}}, \ and\ \bibinfo {author} {\bibfnamefont {E.}~\bibnamefont
  {Tomasi-Gustafsson}},\ }\href {\doibase 10.1016/j.physrep.2014.09.005}
  {\bibfield  {journal} {\bibinfo  {journal} {Phys. Rept.}\ }\textbf {\bibinfo
  {volume} {550-551}},\ \bibinfo {pages} {1} (\bibinfo {year}
  {2015})}\BibitemShut {NoStop}%
\bibitem [{\citenamefont {Pohl}\ \emph {et~al.}(2010)\citenamefont {Pohl} \emph
  {et~al.}}]{Pohl:2010zza}%
  \BibitemOpen
  \bibfield  {author} {\bibinfo {author} {\bibfnamefont {R.}~\bibnamefont
  {Pohl}} \emph {et~al.},\ }\href {\doibase 10.1038/nature09250} {\bibfield
  {journal} {\bibinfo  {journal} {Nature}\ }\textbf {\bibinfo {volume} {466}},\
  \bibinfo {pages} {213} (\bibinfo {year} {2010})}\BibitemShut {NoStop}%
\bibitem [{\citenamefont {Antognini}\ \emph {et~al.}(2013)\citenamefont
  {Antognini} \emph {et~al.}}]{Antognini:1900ns}%
  \BibitemOpen
  \bibfield  {author} {\bibinfo {author} {\bibfnamefont {A.}~\bibnamefont
  {Antognini}} \emph {et~al.},\ }\href {\doibase 10.1126/science.1230016}
  {\bibfield  {journal} {\bibinfo  {journal} {Science}\ }\textbf {\bibinfo
  {volume} {339}},\ \bibinfo {pages} {417} (\bibinfo {year}
  {2013})}\BibitemShut {NoStop}%
\bibitem [{\citenamefont {Mohr}\ \emph {et~al.}(2016)\citenamefont {Mohr},
  \citenamefont {Newell},\ and\ \citenamefont {Taylor}}]{Mohr:2015ccw}%
  \BibitemOpen
  \bibfield  {author} {\bibinfo {author} {\bibfnamefont {P.~J.}\ \bibnamefont
  {Mohr}}, \bibinfo {author} {\bibfnamefont {D.~B.}\ \bibnamefont {Newell}}, \
  and\ \bibinfo {author} {\bibfnamefont {B.~N.}\ \bibnamefont {Taylor}},\
  }\href {\doibase 10.1103/RevModPhys.88.035009} {\bibfield  {journal}
  {\bibinfo  {journal} {Rev. Mod. Phys.}\ }\textbf {\bibinfo {volume} {88}},\
  \bibinfo {pages} {035009} (\bibinfo {year} {2016})},\ \Eprint
  {http://arxiv.org/abs/1507.07956} {arXiv:1507.07956 [physics.atom-ph]}
  \BibitemShut {NoStop}%
\bibitem [{\citenamefont {Pohl}\ \emph {et~al.}(2013)\citenamefont {Pohl},
  \citenamefont {Gilman}, \citenamefont {Miller},\ and\ \citenamefont
  {Pachucki}}]{Pohl:2013yb}%
  \BibitemOpen
  \bibfield  {author} {\bibinfo {author} {\bibfnamefont {R.}~\bibnamefont
  {Pohl}}, \bibinfo {author} {\bibfnamefont {R.}~\bibnamefont {Gilman}},
  \bibinfo {author} {\bibfnamefont {G.~A.}\ \bibnamefont {Miller}}, \ and\
  \bibinfo {author} {\bibfnamefont {K.}~\bibnamefont {Pachucki}},\ }\href
  {\doibase 10.1146/annurev-nucl-102212-170627} {\bibfield  {journal} {\bibinfo
   {journal} {Ann. Rev. Nucl. Part. Sci.}\ }\textbf {\bibinfo {volume} {63}},\
  \bibinfo {pages} {175} (\bibinfo {year} {2013})},\ \Eprint
  {http://arxiv.org/abs/1301.0905} {arXiv:1301.0905 [physics.atom-ph]}
  \BibitemShut {NoStop}%
\bibitem [{\citenamefont {Carlson}(2015)}]{Carlson:2015jba}%
  \BibitemOpen
  \bibfield  {author} {\bibinfo {author} {\bibfnamefont {C.~E.}\ \bibnamefont
  {Carlson}},\ }\href {\doibase 10.1016/j.ppnp.2015.01.002} {\bibfield
  {journal} {\bibinfo  {journal} {Prog. Part. Nucl. Phys.}\ }\textbf {\bibinfo
  {volume} {82}},\ \bibinfo {pages} {59} (\bibinfo {year} {2015})},\ \Eprint
  {http://arxiv.org/abs/1502.05314} {arXiv:1502.05314 [hep-ph]} \BibitemShut
  {NoStop}%
\bibitem [{\citenamefont {Gasparian}(2014)}]{Gasparian:2014rna}%
  \BibitemOpen
  \bibfield  {author} {\bibinfo {author} {\bibfnamefont {A.}~\bibnamefont
  {Gasparian}} (\bibinfo {collaboration} {PRad at JLab}),\ }\bibfield
  {booktitle} {\emph {\bibinfo {booktitle} {{Proceedings, 13th International
  Conference on Meson-Nucleon Physics and the Structure of the Nucleon (MENU
  2013): Rome, Italy, September 30-October 4, 2013}}},\ }\href {\doibase
  10.1051/epjconf/20147307006} {\bibfield  {journal} {\bibinfo  {journal} {EPJ
  Web Conf.}\ }\textbf {\bibinfo {volume} {73}},\ \bibinfo {pages} {07006}
  (\bibinfo {year} {2014})}\BibitemShut {NoStop}%
\bibitem [{\citenamefont {Gilman}\ \emph {et~al.}(2017)\citenamefont {Gilman}
  \emph {et~al.}}]{Gilman:2017hdr}%
  \BibitemOpen
  \bibfield  {author} {\bibinfo {author} {\bibfnamefont {R.}~\bibnamefont
  {Gilman}} \emph {et~al.} (\bibinfo {collaboration} {MUSE}),\ }\href@noop {}
  {\  (\bibinfo {year} {2017})},\ \Eprint {http://arxiv.org/abs/1709.09753}
  {arXiv:1709.09753 [physics.ins-det]} \BibitemShut {NoStop}%
\bibitem [{\citenamefont {Yan}\ \emph {et~al.}(2018)\citenamefont {Yan},
  \citenamefont {Higinbotham}, \citenamefont {Dutta}, \citenamefont {Gao},
  \citenamefont {Gasparian}, \citenamefont {Khandaker}, \citenamefont
  {Liyanage}, \citenamefont {Pasyuk}, \citenamefont {Peng},\ and\ \citenamefont
  {Xiong}}]{Yan:2018bez}%
  \BibitemOpen
  \bibfield  {author} {\bibinfo {author} {\bibfnamefont {X.}~\bibnamefont
  {Yan}}, \bibinfo {author} {\bibfnamefont {D.~W.}\ \bibnamefont
  {Higinbotham}}, \bibinfo {author} {\bibfnamefont {D.}~\bibnamefont {Dutta}},
  \bibinfo {author} {\bibfnamefont {H.}~\bibnamefont {Gao}}, \bibinfo {author}
  {\bibfnamefont {A.}~\bibnamefont {Gasparian}}, \bibinfo {author}
  {\bibfnamefont {M.~A.}\ \bibnamefont {Khandaker}}, \bibinfo {author}
  {\bibfnamefont {N.}~\bibnamefont {Liyanage}}, \bibinfo {author}
  {\bibfnamefont {E.}~\bibnamefont {Pasyuk}}, \bibinfo {author} {\bibfnamefont
  {C.}~\bibnamefont {Peng}}, \ and\ \bibinfo {author} {\bibfnamefont
  {W.}~\bibnamefont {Xiong}},\ }\href {\doibase 10.1103/PhysRevC.98.025204}
  {\bibfield  {journal} {\bibinfo  {journal} {Phys. Rev.}\ }\textbf {\bibinfo
  {volume} {C98}},\ \bibinfo {pages} {025204} (\bibinfo {year} {2018})},\
  \Eprint {http://arxiv.org/abs/1803.01629} {arXiv:1803.01629 [nucl-ex]}
  \BibitemShut {NoStop}%
\bibitem [{\citenamefont {Shmueli}(2010)}]{Shmueli:2010}%
  \BibitemOpen
  \bibfield  {author} {\bibinfo {author} {\bibfnamefont {G.}~\bibnamefont
  {Shmueli}},\ }\href {\doibase 10.1214/10-STS330} {\bibfield  {journal}
  {\bibinfo  {journal} {Statist. Sci.}\ }\textbf {\bibinfo {volume} {25}},\
  \bibinfo {pages} {289} (\bibinfo {year} {2010})}\BibitemShut {NoStop}%
\bibitem [{\citenamefont {Bernauer}\ \emph {et~al.}(2010)\citenamefont
  {Bernauer} \emph {et~al.}}]{Bernauer:2010wm}%
  \BibitemOpen
  \bibfield  {author} {\bibinfo {author} {\bibfnamefont {J.~C.}\ \bibnamefont
  {Bernauer}} \emph {et~al.} (\bibinfo {collaboration} {A1}),\ }\href {\doibase
  10.1103/PhysRevLett.105.242001} {\bibfield  {journal} {\bibinfo  {journal}
  {Phys. Rev. Lett.}\ }\textbf {\bibinfo {volume} {105}},\ \bibinfo {pages}
  {242001} (\bibinfo {year} {2010})},\ \Eprint {http://arxiv.org/abs/1007.5076}
  {arXiv:1007.5076 [nucl-ex]} \BibitemShut {NoStop}%
\bibitem [{\citenamefont {Bernauer}\ \emph {et~al.}(2014)\citenamefont
  {Bernauer} \emph {et~al.}}]{Bernauer:2013tpr}%
  \BibitemOpen
  \bibfield  {author} {\bibinfo {author} {\bibfnamefont {J.~C.}\ \bibnamefont
  {Bernauer}} \emph {et~al.} (\bibinfo {collaboration} {A1}),\ }\href {\doibase
  10.1103/PhysRevC.90.015206} {\bibfield  {journal} {\bibinfo  {journal} {Phys.
  Rev.}\ }\textbf {\bibinfo {volume} {C90}},\ \bibinfo {pages} {015206}
  (\bibinfo {year} {2014})},\ \Eprint {http://arxiv.org/abs/1307.6227}
  {arXiv:1307.6227 [nucl-ex]} \BibitemShut {NoStop}%
\bibitem [{\citenamefont {Lee}\ \emph {et~al.}(2015)\citenamefont {Lee},
  \citenamefont {Arrington},\ and\ \citenamefont {Hill}}]{Lee:2015jqa}%
  \BibitemOpen
  \bibfield  {author} {\bibinfo {author} {\bibfnamefont {G.}~\bibnamefont
  {Lee}}, \bibinfo {author} {\bibfnamefont {J.~R.}\ \bibnamefont {Arrington}},
  \ and\ \bibinfo {author} {\bibfnamefont {R.~J.}\ \bibnamefont {Hill}},\
  }\href {\doibase 10.1103/PhysRevD.92.013013} {\bibfield  {journal} {\bibinfo
  {journal} {Phys. Rev.}\ }\textbf {\bibinfo {volume} {D92}},\ \bibinfo {pages}
  {013013} (\bibinfo {year} {2015})},\ \Eprint
  {http://arxiv.org/abs/1505.01489} {arXiv:1505.01489 [hep-ph]} \BibitemShut
  {NoStop}%
\bibitem [{\citenamefont {Griffioen}\ \emph {et~al.}(2016)\citenamefont
  {Griffioen}, \citenamefont {Carlson},\ and\ \citenamefont
  {Maddox}}]{Griffioen:2015hta}%
  \BibitemOpen
  \bibfield  {author} {\bibinfo {author} {\bibfnamefont {K.}~\bibnamefont
  {Griffioen}}, \bibinfo {author} {\bibfnamefont {C.}~\bibnamefont {Carlson}},
  \ and\ \bibinfo {author} {\bibfnamefont {S.}~\bibnamefont {Maddox}},\ }\href
  {\doibase 10.1103/PhysRevC.93.065207} {\bibfield  {journal} {\bibinfo
  {journal} {Phys. Rev.}\ }\textbf {\bibinfo {volume} {C93}},\ \bibinfo {pages}
  {065207} (\bibinfo {year} {2016})},\ \Eprint
  {http://arxiv.org/abs/1509.06676} {arXiv:1509.06676 [nucl-ex]} \BibitemShut
  {NoStop}%
\bibitem [{\citenamefont {Higinbotham}\ \emph {et~al.}(2016)\citenamefont
  {Higinbotham}, \citenamefont {Kabir}, \citenamefont {Lin}, \citenamefont
  {Meekins}, \citenamefont {Norum},\ and\ \citenamefont
  {Sawatzky}}]{Higinbotham:2015rja}%
  \BibitemOpen
  \bibfield  {author} {\bibinfo {author} {\bibfnamefont {D.~W.}\ \bibnamefont
  {Higinbotham}}, \bibinfo {author} {\bibfnamefont {A.~A.}\ \bibnamefont
  {Kabir}}, \bibinfo {author} {\bibfnamefont {V.}~\bibnamefont {Lin}}, \bibinfo
  {author} {\bibfnamefont {D.}~\bibnamefont {Meekins}}, \bibinfo {author}
  {\bibfnamefont {B.}~\bibnamefont {Norum}}, \ and\ \bibinfo {author}
  {\bibfnamefont {B.}~\bibnamefont {Sawatzky}},\ }\href {\doibase
  10.1103/PhysRevC.93.055207} {\bibfield  {journal} {\bibinfo  {journal} {Phys.
  Rev.}\ }\textbf {\bibinfo {volume} {C93}},\ \bibinfo {pages} {055207}
  (\bibinfo {year} {2016})},\ \Eprint {http://arxiv.org/abs/1510.01293}
  {arXiv:1510.01293 [nucl-ex]} \BibitemShut {NoStop}%
\bibitem [{\citenamefont {Horbatsch}\ \emph {et~al.}(2017)\citenamefont
  {Horbatsch}, \citenamefont {Hessels},\ and\ \citenamefont
  {Pineda}}]{Horbatsch:2016ilr}%
  \BibitemOpen
  \bibfield  {author} {\bibinfo {author} {\bibfnamefont {M.}~\bibnamefont
  {Horbatsch}}, \bibinfo {author} {\bibfnamefont {E.~A.}\ \bibnamefont
  {Hessels}}, \ and\ \bibinfo {author} {\bibfnamefont {A.}~\bibnamefont
  {Pineda}},\ }\href {\doibase 10.1103/PhysRevC.95.035203} {\bibfield
  {journal} {\bibinfo  {journal} {Phys. Rev.}\ }\textbf {\bibinfo {volume}
  {C95}},\ \bibinfo {pages} {035203} (\bibinfo {year} {2017})},\ \Eprint
  {http://arxiv.org/abs/1610.09760} {arXiv:1610.09760 [nucl-th]} \BibitemShut
  {NoStop}%
\bibitem [{\citenamefont {Hayward}\ and\ \citenamefont
  {Griffioen}(2018)}]{Hayward:2018qij}%
  \BibitemOpen
  \bibfield  {author} {\bibinfo {author} {\bibfnamefont {T.~B.}\ \bibnamefont
  {Hayward}}\ and\ \bibinfo {author} {\bibfnamefont {K.~A.}\ \bibnamefont
  {Griffioen}},\ }\href@noop {} {\  (\bibinfo {year} {2018})},\ \Eprint
  {http://arxiv.org/abs/1804.09150} {arXiv:1804.09150 [nucl-ex]} \BibitemShut
  {NoStop}%
\bibitem [{\citenamefont {Higinbotham}\ \emph {et~al.}(2018)\citenamefont
  {Higinbotham}, \citenamefont {Giuliani}, \citenamefont {McClellan},
  \citenamefont {Sirca},\ and\ \citenamefont {Yan}}]{Higinbotham:2018jfh}%
  \BibitemOpen
  \bibfield  {author} {\bibinfo {author} {\bibfnamefont {D.~W.}\ \bibnamefont
  {Higinbotham}}, \bibinfo {author} {\bibfnamefont {P.}~\bibnamefont
  {Giuliani}}, \bibinfo {author} {\bibfnamefont {R.~E.}\ \bibnamefont
  {McClellan}}, \bibinfo {author} {\bibfnamefont {S.}~\bibnamefont {Sirca}}, \
  and\ \bibinfo {author} {\bibfnamefont {X.}~\bibnamefont {Yan}},\ }\href@noop
  {} {\  (\bibinfo {year} {2018})},\ \Eprint {http://arxiv.org/abs/1812.05706}
  {arXiv:1812.05706 [physics.data-an]} \BibitemShut {NoStop}%
\bibitem [{\citenamefont {Hohler}\ \emph {et~al.}(1976)\citenamefont {Hohler},
  \citenamefont {Pietarinen}, \citenamefont {Sabba~Stefanescu}, \citenamefont
  {Borkowski}, \citenamefont {Simon}, \citenamefont {Walther},\ and\
  \citenamefont {Wendling}}]{Hohler:1976ax}%
  \BibitemOpen
  \bibfield  {author} {\bibinfo {author} {\bibfnamefont {G.}~\bibnamefont
  {Hohler}}, \bibinfo {author} {\bibfnamefont {E.}~\bibnamefont {Pietarinen}},
  \bibinfo {author} {\bibfnamefont {I.}~\bibnamefont {Sabba~Stefanescu}},
  \bibinfo {author} {\bibfnamefont {F.}~\bibnamefont {Borkowski}}, \bibinfo
  {author} {\bibfnamefont {G.~G.}\ \bibnamefont {Simon}}, \bibinfo {author}
  {\bibfnamefont {V.~H.}\ \bibnamefont {Walther}}, \ and\ \bibinfo {author}
  {\bibfnamefont {R.~D.}\ \bibnamefont {Wendling}},\ }\href {\doibase
  10.1016/0550-3213(76)90449-1} {\bibfield  {journal} {\bibinfo  {journal}
  {Nucl. Phys.}\ }\textbf {\bibinfo {volume} {B114}},\ \bibinfo {pages} {505}
  (\bibinfo {year} {1976})}\BibitemShut {NoStop}%
\bibitem [{\citenamefont {Belushkin}\ \emph {et~al.}(2007)\citenamefont
  {Belushkin}, \citenamefont {Hammer},\ and\ \citenamefont
  {Meissner}}]{Belushkin:2006qa}%
  \BibitemOpen
  \bibfield  {author} {\bibinfo {author} {\bibfnamefont {M.~A.}\ \bibnamefont
  {Belushkin}}, \bibinfo {author} {\bibfnamefont {H.~W.}\ \bibnamefont
  {Hammer}}, \ and\ \bibinfo {author} {\bibfnamefont {U.~G.}\ \bibnamefont
  {Meissner}},\ }\href {\doibase 10.1103/PhysRevC.75.035202} {\bibfield
  {journal} {\bibinfo  {journal} {Phys. Rev.}\ }\textbf {\bibinfo {volume}
  {C75}},\ \bibinfo {pages} {035202} (\bibinfo {year} {2007})},\ \Eprint
  {http://arxiv.org/abs/hep-ph/0608337} {arXiv:hep-ph/0608337 [hep-ph]}
  \BibitemShut {NoStop}%
\bibitem [{\citenamefont {Lorenz}\ \emph {et~al.}(2012)\citenamefont {Lorenz},
  \citenamefont {Hammer},\ and\ \citenamefont {Meissner}}]{Lorenz:2012tm}%
  \BibitemOpen
  \bibfield  {author} {\bibinfo {author} {\bibfnamefont {I.~T.}\ \bibnamefont
  {Lorenz}}, \bibinfo {author} {\bibfnamefont {H.~W.}\ \bibnamefont {Hammer}},
  \ and\ \bibinfo {author} {\bibfnamefont {U.-G.}\ \bibnamefont {Meissner}},\
  }\href {\doibase 10.1140/epja/i2012-12151-1} {\bibfield  {journal} {\bibinfo
  {journal} {Eur. Phys. J.}\ }\textbf {\bibinfo {volume} {A48}},\ \bibinfo
  {pages} {151} (\bibinfo {year} {2012})},\ \Eprint
  {http://arxiv.org/abs/1205.6628} {arXiv:1205.6628 [hep-ph]} \BibitemShut
  {NoStop}%
\bibitem [{\citenamefont {Alarc\'on}\ and\ \citenamefont
  {Weiss}(2017)}]{Alarcon:2017ivh}%
  \BibitemOpen
  \bibfield  {author} {\bibinfo {author} {\bibfnamefont {J.~M.}\ \bibnamefont
  {Alarc\'on}}\ and\ \bibinfo {author} {\bibfnamefont {C.}~\bibnamefont
  {Weiss}},\ }\href {\doibase 10.1103/PhysRevC.96.055206} {\bibfield  {journal}
  {\bibinfo  {journal} {Phys. Rev.}\ }\textbf {\bibinfo {volume} {C96}},\
  \bibinfo {pages} {055206} (\bibinfo {year} {2017})},\ \Eprint
  {http://arxiv.org/abs/1707.07682} {arXiv:1707.07682 [hep-ph]} \BibitemShut
  {NoStop}%
\bibitem [{\citenamefont {Alarc\'on}\ and\ \citenamefont
  {Weiss}(2018{\natexlab{a}})}]{Alarcon:2017lhg}%
  \BibitemOpen
  \bibfield  {author} {\bibinfo {author} {\bibfnamefont {J.~M.}\ \bibnamefont
  {Alarc\'on}}\ and\ \bibinfo {author} {\bibfnamefont {C.}~\bibnamefont
  {Weiss}},\ }\href {\doibase 10.1103/PhysRevC.97.055203} {\bibfield  {journal}
  {\bibinfo  {journal} {Phys. Rev.}\ }\textbf {\bibinfo {volume} {C97}},\
  \bibinfo {pages} {055203} (\bibinfo {year} {2018}{\natexlab{a}})},\ \Eprint
  {http://arxiv.org/abs/1710.06430} {arXiv:1710.06430 [hep-ph]} \BibitemShut
  {NoStop}%
\bibitem [{\citenamefont {Alarc\'on}\ and\ \citenamefont
  {Weiss}(2018{\natexlab{b}})}]{Alarcon:2018irp}%
  \BibitemOpen
  \bibfield  {author} {\bibinfo {author} {\bibfnamefont {J.~M.}\ \bibnamefont
  {Alarc\'on}}\ and\ \bibinfo {author} {\bibfnamefont {C.}~\bibnamefont
  {Weiss}},\ }\href {\doibase 10.1016/j.physletb.2018.07.060} {\bibfield
  {journal} {\bibinfo  {journal} {Phys. Lett.}\ }\textbf {\bibinfo {volume}
  {B784}},\ \bibinfo {pages} {373} (\bibinfo {year} {2018}{\natexlab{b}})},\
  \Eprint {http://arxiv.org/abs/1803.09748} {arXiv:1803.09748 [hep-ph]}
  \BibitemShut {NoStop}%
\bibitem [{\citenamefont {Ye}\ \emph {et~al.}(2018)\citenamefont {Ye},
  \citenamefont {Arrington}, \citenamefont {Hill},\ and\ \citenamefont
  {Lee}}]{Ye:2017gyb}%
  \BibitemOpen
  \bibfield  {author} {\bibinfo {author} {\bibfnamefont {Z.}~\bibnamefont
  {Ye}}, \bibinfo {author} {\bibfnamefont {J.}~\bibnamefont {Arrington}},
  \bibinfo {author} {\bibfnamefont {R.~J.}\ \bibnamefont {Hill}}, \ and\
  \bibinfo {author} {\bibfnamefont {G.}~\bibnamefont {Lee}},\ }\href {\doibase
  10.1016/j.physletb.2017.11.023} {\bibfield  {journal} {\bibinfo  {journal}
  {Phys. Lett.}\ }\textbf {\bibinfo {volume} {B777}},\ \bibinfo {pages} {8}
  (\bibinfo {year} {2018})},\ \Eprint {http://arxiv.org/abs/1707.09063}
  {arXiv:1707.09063 [nucl-ex]} \BibitemShut {NoStop}%
\bibitem [{\citenamefont {Tanabashi}\ \emph {et~al.}(2018)\citenamefont
  {Tanabashi} \emph {et~al.}}]{Tanabashi:2018oca}%
  \BibitemOpen
  \bibfield  {author} {\bibinfo {author} {\bibfnamefont {M.}~\bibnamefont
  {Tanabashi}} \emph {et~al.},\ }\href {\doibase 10.1103/PhysRevD.98.030001}
  {\bibfield  {journal} {\bibinfo  {journal} {Phys. Rev.}\ }\textbf {\bibinfo
  {volume} {D98}},\ \bibinfo {pages} {030001} (\bibinfo {year}
  {2018})}\BibitemShut {NoStop}%
\bibitem [{\citenamefont {Hill}\ and\ \citenamefont {Paz}(2010)}]{Hill:2010yb}%
  \BibitemOpen
  \bibfield  {author} {\bibinfo {author} {\bibfnamefont {R.~J.}\ \bibnamefont
  {Hill}}\ and\ \bibinfo {author} {\bibfnamefont {G.}~\bibnamefont {Paz}},\
  }\href {\doibase 10.1103/PhysRevD.82.113005} {\bibfield  {journal} {\bibinfo
  {journal} {Phys. Rev.}\ }\textbf {\bibinfo {volume} {D82}},\ \bibinfo {pages}
  {113005} (\bibinfo {year} {2010})},\ \Eprint {http://arxiv.org/abs/1008.4619}
  {arXiv:1008.4619 [hep-ph]} \BibitemShut {NoStop}%
\bibitem [{\citenamefont {Beyer}\ \emph {et~al.}(2017)\citenamefont {Beyer},
  \citenamefont {Maisenbacher}, \citenamefont {Matveev}, \citenamefont {Pohl},
  \citenamefont {Khabarova}, \citenamefont {Grinin}, \citenamefont {Lamour},
  \citenamefont {Yost}, \citenamefont {H{\"a}nsch}, \citenamefont
  {Kolachevsky},\ and\ \citenamefont {Udem}}]{Beyer79}%
  \BibitemOpen
  \bibfield  {author} {\bibinfo {author} {\bibfnamefont {A.}~\bibnamefont
  {Beyer}}, \bibinfo {author} {\bibfnamefont {L.}~\bibnamefont {Maisenbacher}},
  \bibinfo {author} {\bibfnamefont {A.}~\bibnamefont {Matveev}}, \bibinfo
  {author} {\bibfnamefont {R.}~\bibnamefont {Pohl}}, \bibinfo {author}
  {\bibfnamefont {K.}~\bibnamefont {Khabarova}}, \bibinfo {author}
  {\bibfnamefont {A.}~\bibnamefont {Grinin}}, \bibinfo {author} {\bibfnamefont
  {T.}~\bibnamefont {Lamour}}, \bibinfo {author} {\bibfnamefont {D.~C.}\
  \bibnamefont {Yost}}, \bibinfo {author} {\bibfnamefont {T.~W.}\ \bibnamefont
  {H{\"a}nsch}}, \bibinfo {author} {\bibfnamefont {N.}~\bibnamefont
  {Kolachevsky}}, \ and\ \bibinfo {author} {\bibfnamefont {T.}~\bibnamefont
  {Udem}},\ }\href {\doibase 10.1126/science.aah6677} {\bibfield  {journal}
  {\bibinfo  {journal} {Science}\ }\textbf {\bibinfo {volume} {358}},\ \bibinfo
  {pages} {79} (\bibinfo {year} {2017})}\BibitemShut {NoStop}%
\bibitem [{\citenamefont {Fleurbaey}\ \emph {et~al.}(2018)\citenamefont
  {Fleurbaey}, \citenamefont {Galtier}, \citenamefont {Thomas}, \citenamefont
  {Bonnaud}, \citenamefont {Julien}, \citenamefont {Biraben}, \citenamefont
  {Nez}, \citenamefont {Abgrall},\ and\ \citenamefont
  {Guéna}}]{Fleurbaey:2018fih}%
  \BibitemOpen
  \bibfield  {author} {\bibinfo {author} {\bibfnamefont {H.}~\bibnamefont
  {Fleurbaey}}, \bibinfo {author} {\bibfnamefont {S.}~\bibnamefont {Galtier}},
  \bibinfo {author} {\bibfnamefont {S.}~\bibnamefont {Thomas}}, \bibinfo
  {author} {\bibfnamefont {M.}~\bibnamefont {Bonnaud}}, \bibinfo {author}
  {\bibfnamefont {L.}~\bibnamefont {Julien}}, \bibinfo {author} {\bibfnamefont
  {F.}~\bibnamefont {Biraben}}, \bibinfo {author} {\bibfnamefont
  {F.}~\bibnamefont {Nez}}, \bibinfo {author} {\bibfnamefont {M.}~\bibnamefont
  {Abgrall}}, \ and\ \bibinfo {author} {\bibfnamefont {J.}~\bibnamefont
  {Guéna}},\ }\href {\doibase 10.1103/PhysRevLett.120.183001} {\bibfield
  {journal} {\bibinfo  {journal} {Phys. Rev. Lett.}\ }\textbf {\bibinfo
  {volume} {120}},\ \bibinfo {pages} {183001} (\bibinfo {year} {2018})},\
  \Eprint {http://arxiv.org/abs/1801.08816} {arXiv:1801.08816
  [physics.atom-ph]} \BibitemShut {NoStop}%
\bibitem [{\citenamefont {Peset}\ and\ \citenamefont
  {Pineda}(2014)}]{Peset:2014jxa}%
  \BibitemOpen
  \bibfield  {author} {\bibinfo {author} {\bibfnamefont {C.}~\bibnamefont
  {Peset}}\ and\ \bibinfo {author} {\bibfnamefont {A.}~\bibnamefont {Pineda}},\
  }\href {\doibase 10.1016/j.nuclphysb.2014.07.027} {\bibfield  {journal}
  {\bibinfo  {journal} {Nucl. Phys.}\ }\textbf {\bibinfo {volume} {B887}},\
  \bibinfo {pages} {69} (\bibinfo {year} {2014})},\ \Eprint
  {http://arxiv.org/abs/1406.4524} {arXiv:1406.4524 [hep-ph]} \BibitemShut
  {NoStop}%
\bibitem [{\citenamefont {Hoferichter}\ \emph {et~al.}(2016)\citenamefont
  {Hoferichter}, \citenamefont {Kubis}, \citenamefont {de~Elvira},
  \citenamefont {Hammer},\ and\ \citenamefont
  {Meissner}}]{Hoferichter:2016duk}%
  \BibitemOpen
  \bibfield  {author} {\bibinfo {author} {\bibfnamefont {M.}~\bibnamefont
  {Hoferichter}}, \bibinfo {author} {\bibfnamefont {B.}~\bibnamefont {Kubis}},
  \bibinfo {author} {\bibfnamefont {J.~R.}\ \bibnamefont {de~Elvira}}, \bibinfo
  {author} {\bibfnamefont {H.-W.}\ \bibnamefont {Hammer}}, \ and\ \bibinfo
  {author} {\bibfnamefont {U.-G.}\ \bibnamefont {Meissner}},\ }\href {\doibase
  doi:10.1140/epja/i2016-16331-7} {\bibfield  {journal} {\bibinfo  {journal}
  {Eur.\ Phys.\ J.\ A}\ }\textbf {\bibinfo {volume} {52}},\ \bibinfo {pages}
  {331} (\bibinfo {year} {2016})},\ \Eprint {http://arxiv.org/abs/1609.06722}
  {arXiv:1609.06722 [hep-ph]} \BibitemShut {NoStop}%
\end{thebibliography}%
\end{document}